\documentclass[prl,twocolumn,a4paper,amsmath,floatfix]{revtex4}
\usepackage[utf8]{inputenc}
\usepackage{graphicx}
\usepackage{amsmath}

\usepackage{color}
\newcommand{\WS}{Wahnstr\"om}
\newcommand{\KA}{Kob-Andersen}

\begin{document}

\title{Autonomously revealing hidden local structures in supercooled liquids}

\author{Emanuele Boattini$^1$,  Susana Mar\'in-Aguilar$^2$, Saheli Mitra$^2$, Giuseppe Foffi$^2$, Frank Smallenburg$^2$, Laura Filion$^1$}
\affiliation{$^1$Soft Condensed Matter, Debye Institute of Nanomaterials Science, Utrecht Utrecht, Netherlands \\
$^2$Universit\'e Paris-Saclay, CNRS, Laboratoire de Physique des Solides, 91405 Orsay, France}

\maketitle

{\bf Few questions in condensed matter science have proven as difficult to unravel as the interplay between structure and dynamics in supercooled liquids and glasses. The conundrum: close to the glass transition, the dynamics slow down dramatically and become heterogeneous \cite{ediger2000spatially, berthier2011dynamical} while the structure appears largely unperturbed. Largely unperturbed, however, is not the same as unperturbed, and many studies have attempted to identify ``slow'' local structures by exploiting dynamical information \cite{royall2015role, tanaka2019revealing}. Nonetheless, the question remains open: is the key to the slow dynamics imprinted in purely structural information? And if so, is there a way to determine the relevant structures without any dynamical information?  Here, we use a newly developed unsupervised machine learning (UML) algorithm to  identify structural heterogeneities in three archetypical glass formers. In each system, the UML approach autonomously designs an order parameter based purely on structural variation within a single snapshot. Impressively, this order parameter strongly correlates with the dynamical heterogeneity. Moreover, the structural characteristics linked to slow particles disappear as we move away from the glass transition. Our results demonstrate the power of machine learning techniques to detect structural patterns even in disordered systems, and provide a new way forward for unraveling the structural origins of the slow dynamics of glassy materials.}

Machine learning (ML) techniques are rapidly becoming a game-changer in the study of materials. Examples include speeding up computationally expensive calculations \cite{behler2007generalized}, accurately distinguishing different crystal phases \cite{geiger2013neural,boattini2018neural}, and even developing design rules for structural and material properties \cite{butler2018machine}. 
An exciting and intriguing development is the design of unsupervised machine learning (UML) algorithms that can autonomously classify particles based on patterns in their local environment \cite{reinhart2017machine,boattini2019unsupervised,spellings2018machine}. A key strength of these UML approaches is that they can find variations in local structure without any {\em a priori} knowledge of what might appear, opening the door to finding new, unanticipated structures. 

The idea of an autonomous algorithm that picks out structural heterogeneities is a particularly appealing one in the context of supercooled liquids. In this field, the last few years have seen a frantic hunt for local structural features that can be interpreted as the underlying cause for dynamical heterogeneities. To this end, a number of studies have examined the prevalence and lifetimes of a large variety of locally favored structures \cite{malins2013identification, leocmach2012roles}, correlated dynamics with local order parameters based on e.g. tetrahedrality or packing efficiency \cite{marin2019tetrahedrality, tong2018revealing, tong2019structural}, and have looked at the dynamical effects of promoting specific local features \cite{taffs2016role, marin2019slowing, speck2012first}. In an impressive application of supervised machine learning techniques, it was shown that support-vector machines could be taught to recognize more mobile particles in several glass formers \cite{cubuk2015identifying,  schoenholz2016structural}. However, in order to train them,  data linking structure to future dynamics had to be used. Overall, these studies demonstrate that local structure indeed provides strong clues for the local dynamics of a given region, but the best way to look for these local structures depends strongly on the system under consideration. It would be of great help to devise a method that, based {\it solely} on the real space static structure, can autonomously detect structural heterogeneities. Here, UML techniques offer a novel and unbiased fresh look at the problem.

\begin{figure*}
\includegraphics[width=0.9\linewidth]{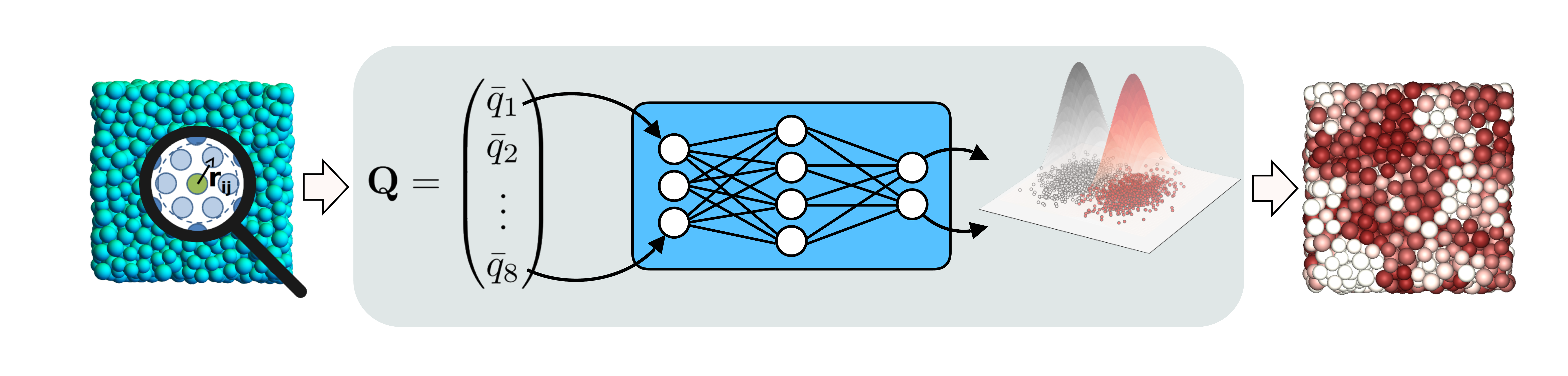}
\caption{{\bf Schematic representation of the unsupervised machine learning method:} In this method, the local environment of a particle is encoded in a vector ($\mathbf Q$) of bond order parameters, which is used as the input for an artificial neural network trained to reduce its dimensionality. The resulting distribution of particle environments in the lower dimension is clustered using a Gaussian mixture model. Finally, particles are assigned a probability of belonging to one of the two clusters, and colored accordingly.} \label{fig:method}
\end{figure*}

\begin{figure*}[t]

\includegraphics[width=0.9\linewidth]{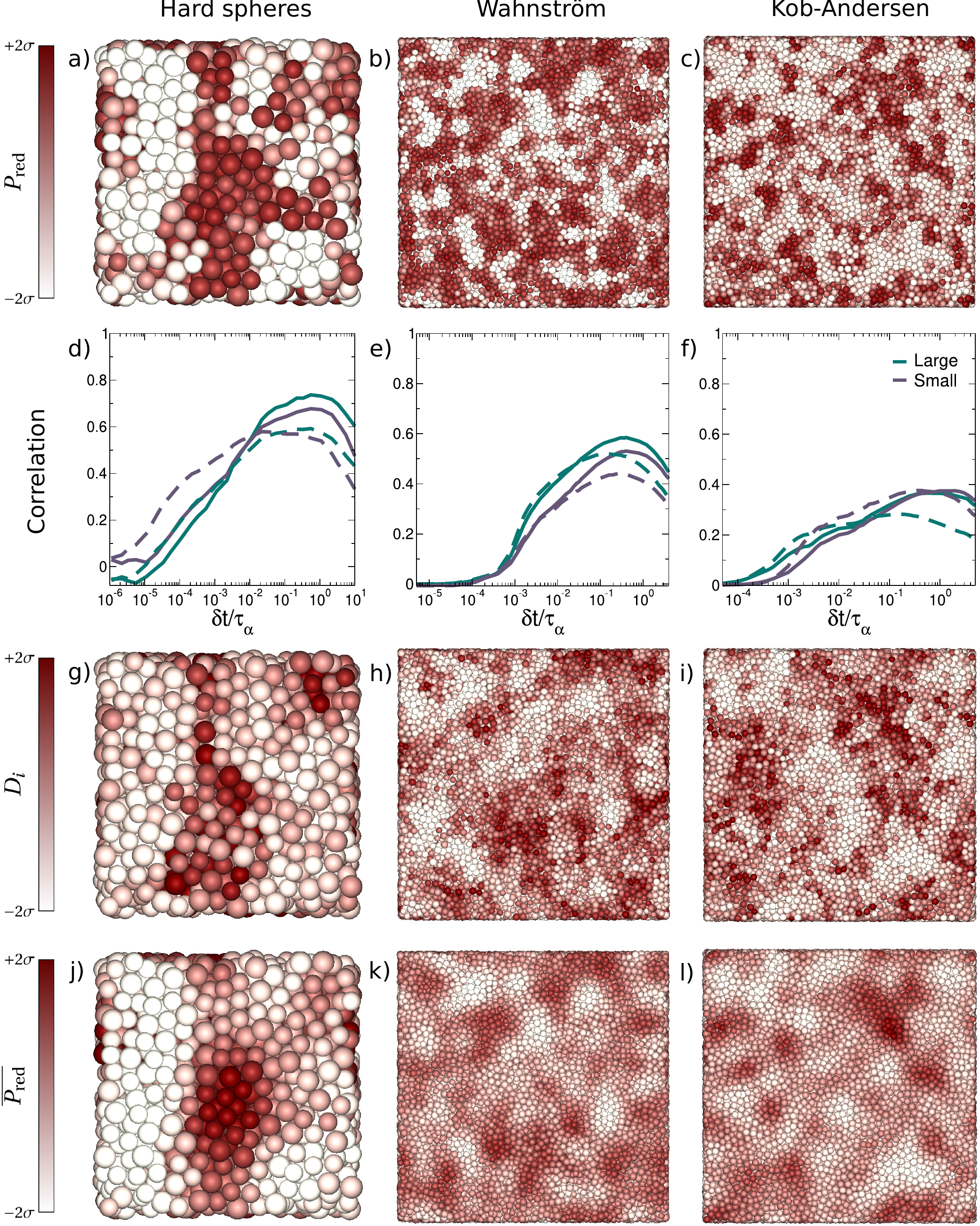}
\caption{{\bf Structural analysis and correlations with dynamics in three archetypical glass formers:}\\ {\bf (a-c)} Snapshots of different glassy models. From left to right: hard spheres with packing fraction $\eta=0.58$, size ratio $q=0.85$ and composition $x_L=0.3$, Wahnstr\"om at density $\rho^*=0.81$ and temperature $T^*=0.7$, and Kob-Andersen at density $\rho^*=1.2$ and temperature $T^*=0.5$. Particles are colored according to their membership probability $P_\mathrm{red}$ of belonging to a specific cluster identified by the machine learning approach. In particular, particles whose $P_\mathrm{red}$ is two or more standard deviations $\sigma$ above the mean value are dark red, while particles with $P\mathrm{red}$ more than two $\sigma$ below the mean are colored white. {\bf (d-f)} Spearman's rank correlation between the particles' dynamic propensity $D_i$ and either their membership probability $P_{\text{red}}(i)$ (dashed lines) or its local average, $\bar{P}_{\text{red}}(i)$, over a local spherical neighbourhood with radius of two times the diameter of the large spheres (solid lines). {\bf (g-i)} Same snapshots as (a-c), but colored according to the dynamic propensity $D_i$. {\bf (j-l)} Same snapshots, colored according to the locally averaged membership probability  $\bar{P}_{\text{red}}(i)$.} \label{fig:corr}
\end{figure*}

To explore whether UML techniques can indeed be harnessed to detect structural heterogeneities, we examine the structure of three archetypical glass forming systems: binary hard spheres, Wahnstr\"om , and Kob-Andersen, described in detail in the Methods. These three model systems have been extensively studied in the context of fundamental glass formers, and have proven extremely valuable in unraveling many aspects of the glass transition (see e.g. \cite{royall2015role}). As such, these models provide an ideal playground for testing the ability of UML techniques to find local structural features in supercooled liquids.



Over the last few years, a number of unsupervised machine learning techniques for classifying local structure have been proposed \cite{reinhart2017machine,spellings2018machine,boattini2019unsupervised}, using different definitions of local structures, and different approaches for classification. Here, we use a method that we recently developed \cite{boattini2019unsupervised} for detecting crystalline structure. In this approach, the local environment of each particle is described as a vector of eight bond order parameters (see Methods). We then use an auto-encoder to lower the dimensionality of this vector (see Supplemental Information (SI)). The auto-encoder is a neural network trained to reproduce its input as its output. This neural network is especially designed to contain a ``bottleneck'' with a lower dimensionality than the input vector, such that the network is forced to compress the information, and subsequently decompress it again. After training the auto-encoder, we only retain the compression part of the network, and use it as our dimensionality reducer. The particles are then grouped in this lower dimensional space into two clusters using Gaussian mixture models. Based on this clustering, each particle is assigned a probability to belong to one of the two clusters, e.g. $P_\mathrm{red}$ for the cluster (arbitrarily) labeled as red. This probability can then be interpreted as an order parameter describing the largest structural heterogeneities in the system, as found by the UML approach. A schematic of this classification method is shown in Fig. \ref{fig:method}, and described in detail in Ref. \cite{boattini2019unsupervised} and the SI.
We would like to stress that in our approach no dynamical information is used in the training process, in contrast with previous supervised ML studies of glasses \cite{cubuk2015identifying, schoenholz2016structural}. In fact, our auto-encoder is trained on a {\it  single static snapshot} for each system. 

To begin our investigation, we select one equilibrated configuration in the glassy regime for each glass former and perform the UML analysis. In Figure \ref{fig:corr}a-c, we show the results of the UML analysis on each snapshot, by coloring each particle according to the probability they belong to the ``red'' cluster. Using this order parameter, the system shows clear structural heterogeneity, consisting of regions of both environments.  

The question now is whether these environments are correlated to the dynamics.  To probe this, we measure the dynamic propensity  $D_i(\delta t)$ of particle $i$: a measure for how mobile particle $i$ will be over the next time interval $\delta t$ (see Methods), which has proven useful in supercooled liquids \cite{berthier2007structure, widmer2006predicting, marin2019tetrahedrality, tong2018revealing}. In Fig. \ref{fig:corr}d-f, we plot the Spearman's correlation coefficient between $P_\mathrm{red}$ and $D_i(\delta t)$, as a function of the time interval $\delta t$. As one might expect, this correlation is weak both for very short time scales, where particles are simply rattling within their cages, and for long time scales where the system loses its memory of the initial configurations.  It peaks slightly below the structural relaxation time  $\tau_\alpha$ indicating that we have indeed identified structures connected to the structural relaxation.

\begin{figure*}
\center
\begin{tabular}{lllll}
    \includegraphics[width=0.3\linewidth]{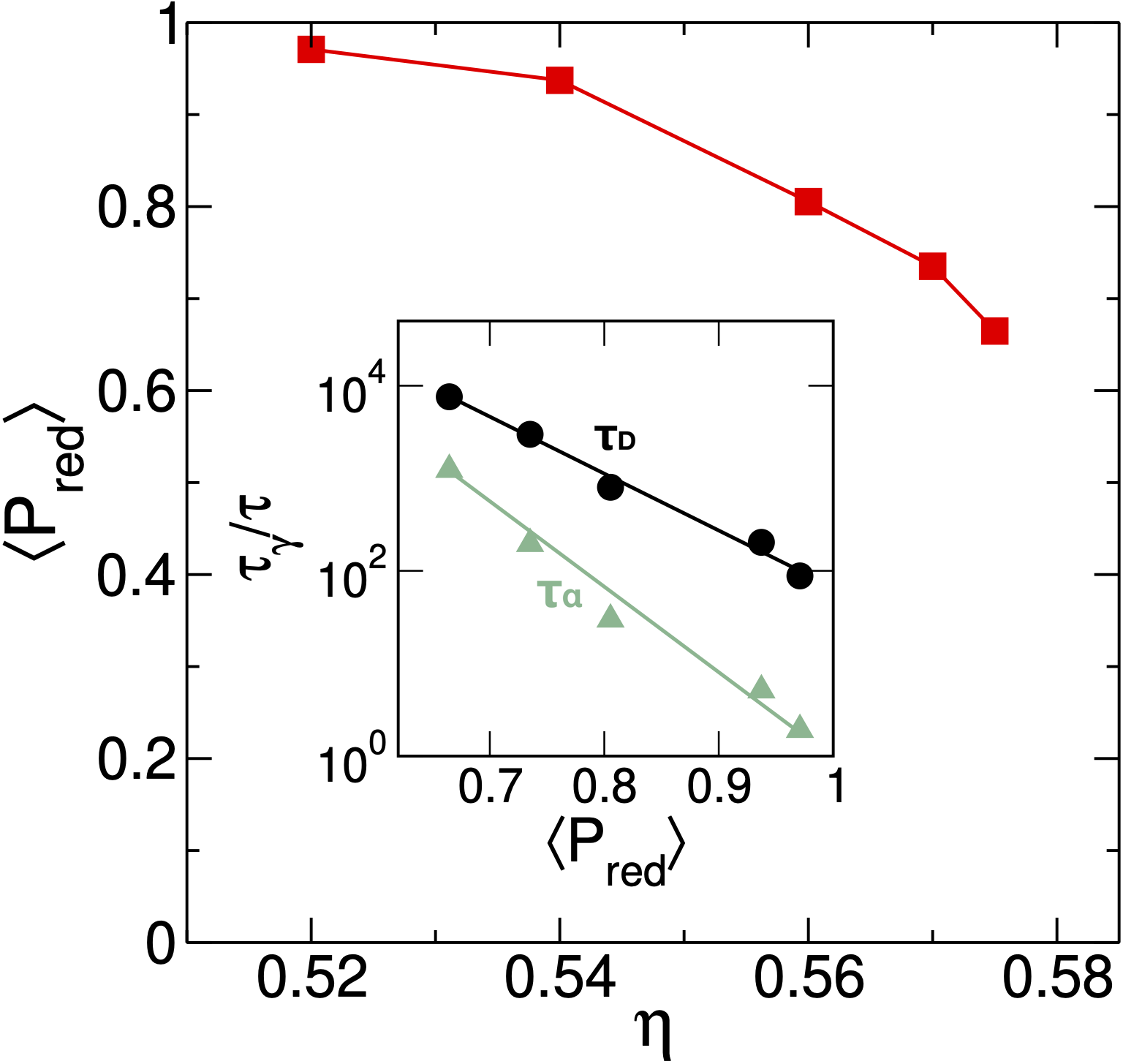}
&& \includegraphics[width=0.3\linewidth]{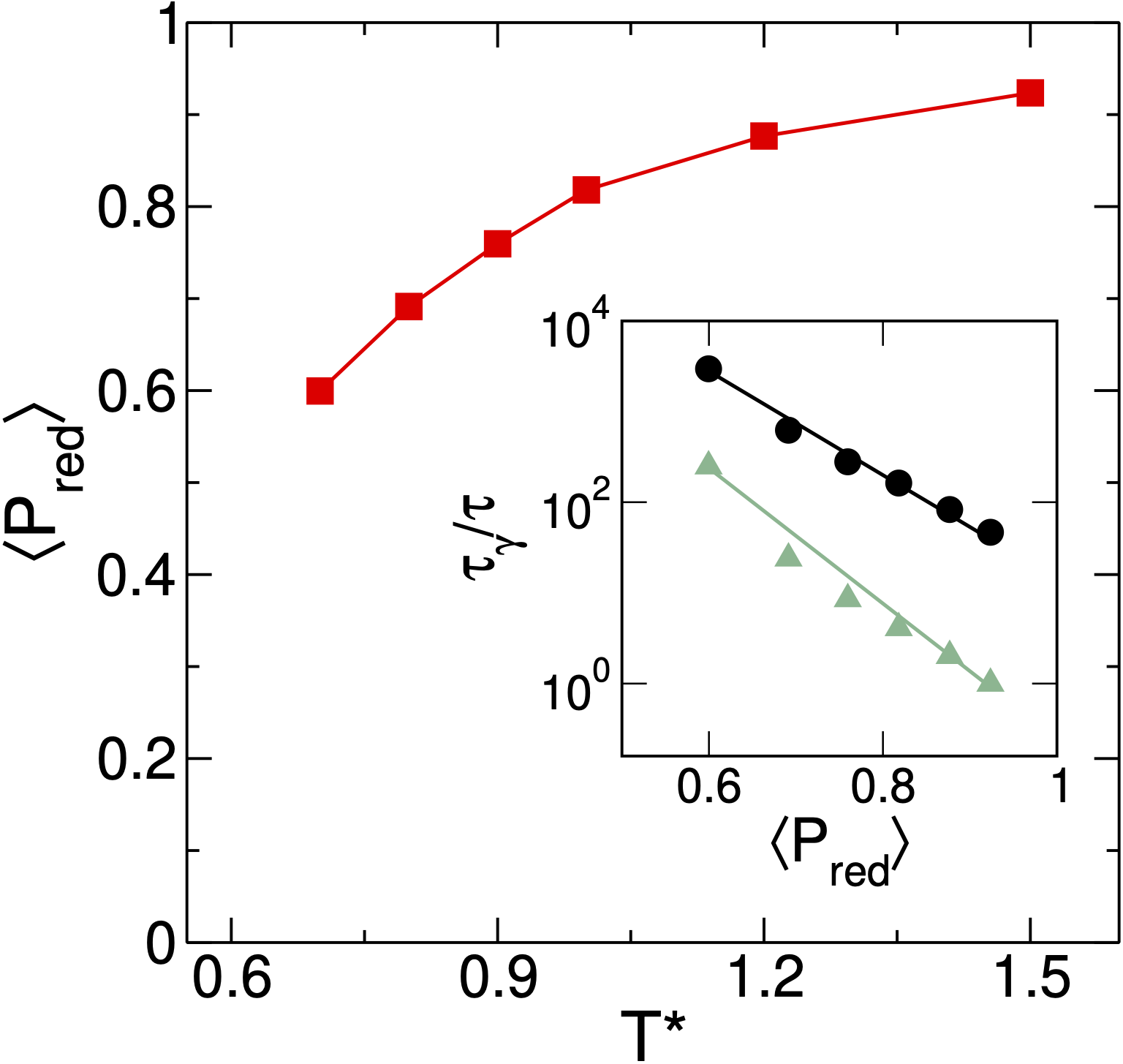}
&& \includegraphics[width=0.3\linewidth]{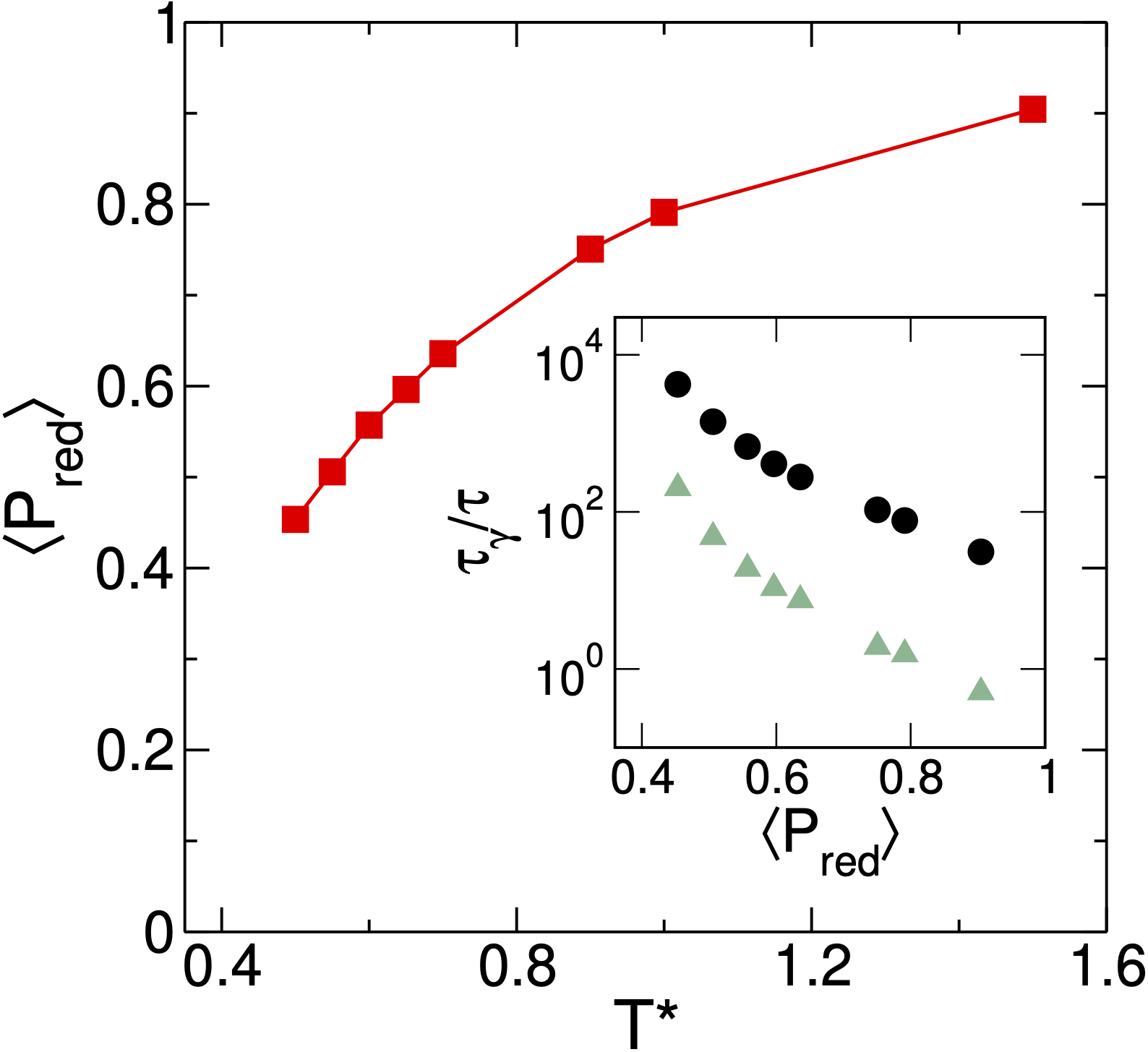}
\vspace{-5cm} \\a)&&b)&&c)
\end{tabular}
\vspace{4.75cm}
\caption{{\bf Structural order parameter as a function of supercooling:} Mean membership probability $\left<P_{\text{red}}\right>$ for the three systems: (a) as a function of the packing fraction $\eta$ for hard spheres, and as a function of the reduced temperature $T^*$ for (b) Wahnstr\"{o}m and (c) Kob-Andersen. The insets show the relation between $\left<P_{\text{red}}\right>$  and either the structural relaxation time $\tau_\alpha$ (green triangles), or the diffusion time $\tau_D$ (black circles). In all cases, $P_\text{red}$ is calculated using the UML approach trained at the highest degree of supercooling.} \label{fig:resultsOP}
\end{figure*}

\begin{figure*}
\center
\begin{tabular}{lllll}
\includegraphics[width=0.3\linewidth]{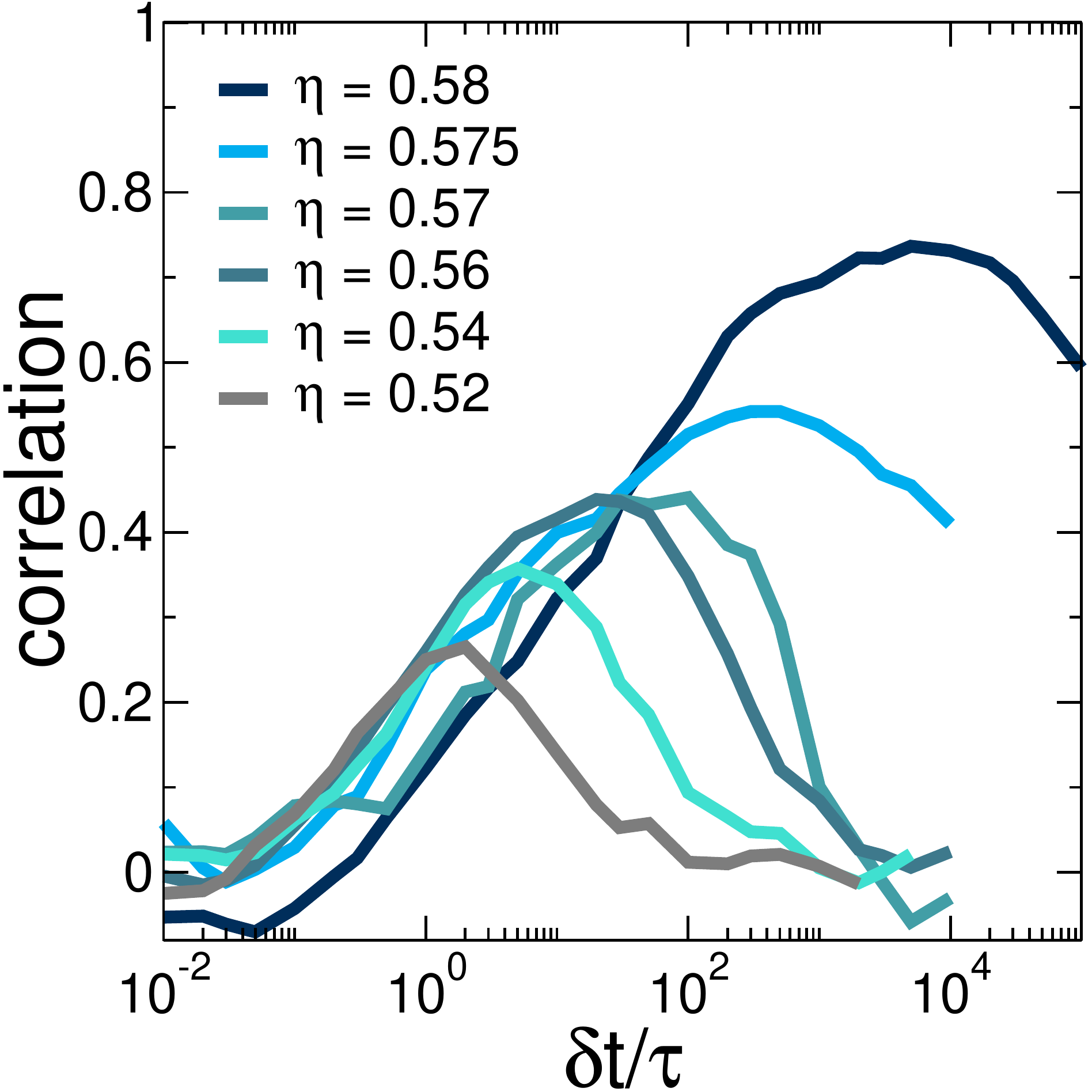}&&
\includegraphics[width=0.3\linewidth]{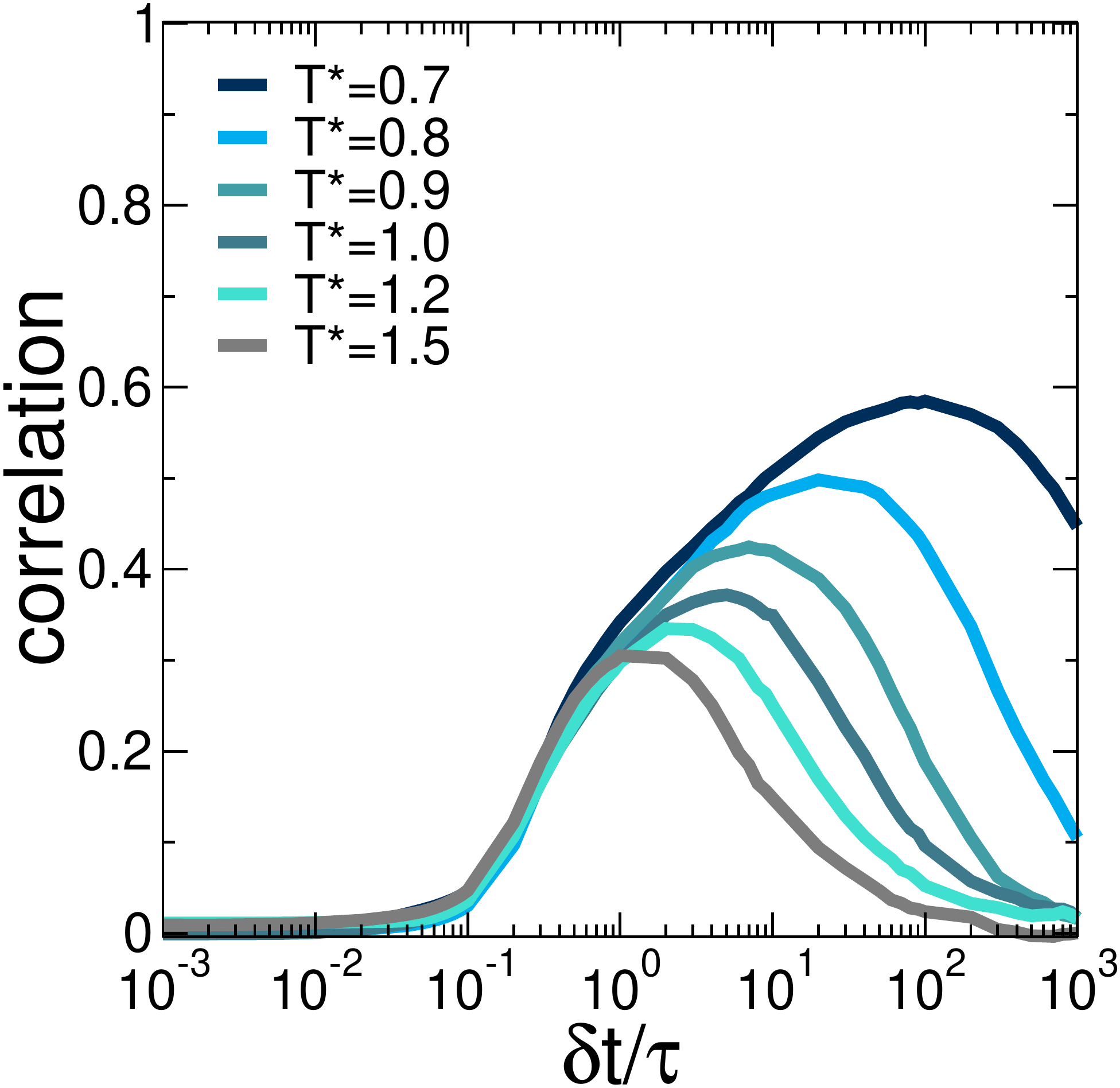}&&
\includegraphics[width=0.3\linewidth]{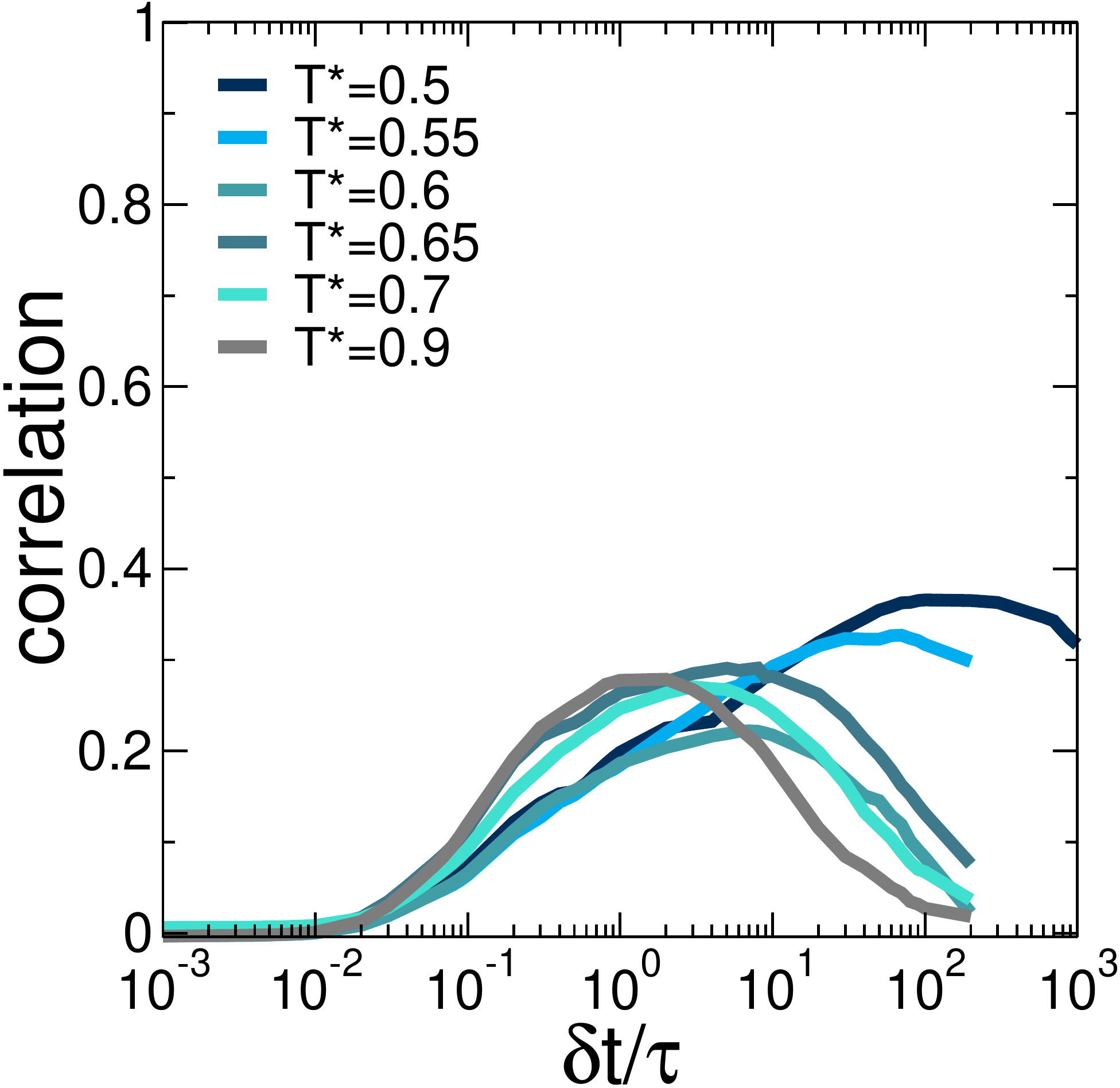}\vspace{-5.25cm}\\
a) && b)&& c)
\end{tabular}
\vspace{5cm}
\caption{{\bf Correlation between structure and dynamics for different supercoolings:} Correlation between the locally averaged $\bar{P}\mathrm{red}$ and dynamic propensity for large particles (see SI for small particles). Note that the averaging radius for $\bar{P}_\mathrm{red}$ is $2 d_L$ in all cases. In all cases, the UML approach is retrained on a snapshot from the system in question.} \label{fig:correlationsT}
\end{figure*}

To further investigate the correlation between the UML classification and the dynamics, in Fig. \ref{fig:corr}g-i, we color the particles according to their dynamic propensity, with $\delta t$ chosen to correspond to the maximum in the correlation. Clearly, regions of high dynamic propensity correspond to high values of $P_\mathrm{red}$, indicating that the particles identified as part of the red cluster also largely correspond to the faster particles in the system.  The correlation can be further improved by averaging $P_\mathrm{red}$ over particles within a small local region, as demonstrated by both the solid lines in d-f and the snapshots in j-l. In all cases, the correlation between the averaged $\bar{P}_\mathrm{red}$ and $D_i$ peaks very close to $\tau_{\alpha}$. This is slightly later than the unaveraged version, likely because we are now looking at larger regions, which will take more time to rearrange.

To summarize, in all cases the UML approach has identified  an order parameter which captures both the local and global dynamics using, as a training set, only static local structure information. Interestingly, this order parameter performs approximately as well as or better than many previously introduced order parameters \cite{marin2019tetrahedrality, hocky2014correlation, tong2019structural, paret2020assessing}.
As also found in previous work \cite{hocky2014correlation}, \KA{} seems to be the model whose behaviour is less captured by our analysis. This might be related to the attraction that could induce heterogeneities over large length scales due to the proximity of a gas-liquid phase coexistence \cite{sastry2000liquid, berthier2009nonperturbative}. This kind of effect would not be fully captured by our (highly local) observables.  

The natural next question is whether one of the structural groups detected by the UML approach becomes more dominant as we move away from the glass transition. To this end, we use the exact same UML order parameter trained on the snapshots of Fig. \ref{fig:corr} on systems equilibrated at lower degrees of supercooling: lower packing fractions $\eta$ for hard spheres, and higher temperatures $T^*$ for the other two models. In Fig. \ref{fig:resultsOP}, we plot the average value of $P_\mathrm{red}$ as a function of the degree of supercooling for each glass former.  In all cases, $P_\mathrm{red}$ increases monotonically as the system moves out of the glassy regime. Hence, the structures we identify as white (slow) at strong supercooling, disappear as we move away from the glass transition -- clearly showing that the UML order parameter identifies local structures that are important for the dynamical slowdown.
Interestingly, as shown in the insets in Fig. \ref{fig:resultsOP}, the relationship between $P_\mathrm{red}$ and the structural relaxation time is exponential for both the hard-sphere and the Wahnstr\"om system. 

Finally, the dynamics should become less heterogeneous (and hence less predictable) as we move away from the glass transition. To test this, we perform a new UML analysis on the glass formers at different packing fractions and temperatures. Specifically, for each state point we find a new projection and classification, and determine the correlation between $P_\mathrm{red}$ and $D_i$.  In Fig. \ref{fig:correlationsT} we show that indeed, the correlations become weaker and shift to shorter times (along with $\tau_\alpha$) as we move away from the glassy regime. This further confirms that the UML algorithm correctly identifies the local structures that are important for dynamical heterogeneity.

Clearly, the UML classifies particles into two groups that turn out to have very different dynamics. So what is the structural difference between these groups? As the UML is based on bond order parameters, a natural first check is to examine the differences in bond order between the groups. Perhaps surprisingly, the average bond order parameters for particles in each group do not show dramatic differences -- with the most remarkable observation being that the fast particles correspond to higher overall bond order (see SI). 
As the BOPs do not show a dramatic difference, we explore another avenue for differentiating the local structure in each group:  topological cluster classification (TCC) \cite{malins2013tcc}. This algorithm detects a set of pre-defined clusters in each system. We find that for the hard sphere and \WS{} systems, the slow cluster correlates strongly with local structures built up out of one or more tetrahedra, while membership of the fast cluster is correlated with TCC clusters built from square pyramids. For the \KA{} mixture, the slow cluster still correlates best with tetrahedral environments, but correlations are significantly weaker. Interestingly, TCC detects essentially no clusters that correlate with the fast cluster, suggesting that these particles have local environments not detected by TCC. This is one area where the UML approach shines: it is not restricted by {\em a priori} assumptions about the features that are considered in the clustering. 

All our results are consistent with the picture of these supercooled liquids consisting of two competing structural populations \cite{speck2012first}. As the system is pushed closer to the glass transition, one group becomes more dominant, due to a more favorable local packing or potential energy \cite{royall2015role}. Intuitively, this more stable group is also less mobile, and hence its emergence has a profound impact on both local and global dynamics. Note that a similar two-state picture has been extremely successful in understanding the glassy behavior of supercooled water \cite{shi2018origin, caupin2019thermodynamics}, where the competing local structures are ostensibly linked to different thermodynamic phases at extreme supercooling. 

In conclusion, we have demonstrated that a simple and fast auto-encoder-based UML approach is a powerful tool in the development of new {\it structural} order parameters in supercooled liquids. Impressively, the structural heterogeneities captured by this order parameter turn out to be strongly correlated to the dynamical heterogeneities in all three glass formers studied here, creating a new way forward for unraveling the microscopic origins of dynamical slowdown in supercooled liquids.

\section{Acknowledgements}
We gratefully acknowledge Alfons van Blaaderen for many useful discussions. L.F. and E.B. gratefully acknowledge funding from The Netherlands  Organisation  for  Scientific  Research  (NWO)  (Grant  No.16DDS004), and L.F. acknowledges funding from NWO for a Vidi grant. 
S.M.-A. acknowledges support from the Consejo Nacional de Ciencia y Tecnología (CONACyT scholarship No. 340015/471710).

\section{Author Contributions}
L.F and F.S. together suggested the study.  L.F., F.S, G.F, E.B. and S.M.-A designed the research.  S.M.-A. and S.M. preformed the simulations, and E.B. and S.M.-A. preformed the data analysis.  All authors contributed to the interpretation of the results and wrote the paper.

\section{Methods}
\footnotesize

\subsection{Models}

We consider three model glass formers: binary hard spheres, \WS{}\cite{wahnstrom1991molecular}, and \KA{}\cite{kob1995testing}. Both \WS{} and \KA{} are binary mixtures of Lennard-Jones (LJ) particles. 

The binary hard-sphere model we consider is a mixtures of 30\% large $A$ particles and 70\% small $B$ particles, with size ratio $\sigma_B/\sigma_A = 0.85$. 

The \WS{} model \cite{wahnstrom1991molecular} is an equimolar (50\%-50\%) mixture of $A$ and $B$ particles. The LJ interaction strength between all pairs of particles is identical ($\epsilon_{AA} = \epsilon_{AB} = \epsilon_{BB}$), but the $B$ particles are slightly larger than the $A$ particles ($\sigma_{BB} = 1.2 \sigma_{AA}$ and $\sigma_{AB} = 1.1 \sigma_{AA}$). The LJ potential is truncated and shifted at the minimum in the potential, such that the interactions are purely repulsive.

The \KA{} model \cite{kob1995testing} is a non-additive mixture of 80\% (large) $A$ particles and 20\% (small) $B$ particles. The interaction parameters are $\sigma_{BB} = 0.88 \sigma_{AA}$, $\sigma_{AB} = 0.8 \sigma_{AA}$, $\epsilon_{BB} = 0.5 \epsilon_{AA}$, and $\epsilon_{AB} = 1.5 \epsilon_{AA}$. The LJ potential is truncated and shifted at a cutoff distance $r_{c,ij} = 2.5 \sigma_{ij}$ (where $i,j \in \{A,B\}$), such that the attractive part of the potential is retained. 

For both \WS{} and \KA{}, we define the reduced number density $\rho^* = \rho \sigma_{AA}^3$ and reduced temperature $T^* = k_B T / \epsilon_{AA}$, with $k_B$ Boltzmann's constant.

\subsection{Simulations}

For all models, we use molecular dynamics simulations in the canonical ensemble. In the case of hard spheres, the simulations are performed using an event-driven approach. For \WS{} and \KA{}, we use the simulation package LAMMPS \cite{plimpton1995fast}. 

Dynamic propensities are calculated as an isoconfigurational ensemble average of the absolute  displacement of each particle. In other words, we perform at least 32 independent simulations starting from the same initial configuration, but with randomly chosen velocities for all particles. The dynamic propensity of particle $i$ after a time interval $\delta t$ is then defined as
\begin{equation}
D_i(\delta t) = \langle \left|\mathbf{r_i(\delta t)} - \mathbf{r_i(0)}\right|\rangle_\mathrm{c},
\end{equation}
where $\mathbf{r}_i(t)$ is the position of particle $i$ at time $t$, and the average is taken over the independent runs.

In order to obtain the relaxation time $\tau_\alpha$, we calculate the self-intermediate scattering function (ISF) for the \WS{} and the \KA{} systems, and the total intermediate scattering function for the hard spheres:
\begin{equation}
    F(q,t)=\frac{\left<\sum_{j,k}\exp\left\{ i\mathbf{q}\left[\mathbf{r}_j(t)-\mathbf{r}_k(0)\right]\right\}\right>}{\left< \sum_{j,k}\exp \left\{ i\mathbf{q}\left[\mathbf{r}_j(0)-\mathbf{r}_k(0)\right]\right\}\right>},
\end{equation}
where $\mathbf{r}_{i}$ is the position of particle $i$ and $\mathbf{q}$ is a wave vector. We calculate the ISF at an inverse wavelength $q = |\mathbf{q}|$ corresponding to the first peak of the general structure factor. After that, we fit the long-time decay of the ISF with a stretched exponential function $\gamma \exp \left[ - ( t/\tau_\alpha)^\beta \right]$, where $\gamma$, $\beta$ and the relaxation time $\tau_\alpha$ are fit parameters.

\subsection{Local environment description}
To characterize the local environment of each particle, we use an averaged version of the local bond order parameters (BOPs) introduced by Steinhardt \emph{et al.} \cite{steinhardt1983bond}. First, we define for any given particle $i$ the complex quantities
\begin{equation}
\label{qlm}
q_{lm}(i) = \frac{1}{N_b(i)}\sum_{j \in \mathcal{N}_b(i)} Y^m_l(\mathbf{r}_{ij}),
\end{equation}
where $Y^m_l(\mathbf{r_{ij}})$ are the spherical harmonics of order $l$, with $m$ an integer that runs from $m=-l$ to $m=+l$. Additionally, $\mathbf{r}_{ij}$ is the vector from particle $i$ to particle $j$, and $\mathcal{N}_b(i)$ is the set of nearest neighbors of particle $i$, which we will define later.  Note that $\mathcal{N}_b(i)$ contains $N_b(i)$ particles.  Then, the rotationally invariant BOPs, $q_l$, are defined as \cite{steinhardt1983bond}
\begin{equation}
\label{ql}
q_l(i) = \sqrt{\frac{4\pi}{2l+1}\sum_{m=-l}^{l}| q_{lm}(i)|^2}.
\end{equation}
Finally, we define an average $\bar{q}_l(i)$ as
\begin{equation}
\label{avql}
\bar{q}_{l}(i) = \frac{1}{{N}_b(i) +1}\left[q_l(i)+\sum_{k\in \mathcal{N}_b(i)} q_{l}(k)\right].
\end{equation}
Note that the quantities in Eq. \ref{avql} differ from the averaged BOPs introduced by Lechner and Dellago \cite{Lechner2008} where first the averaging is performed on the non-rotational invariant $q_{lm}$, and then rotational-invariant quantities are built.   

Our description of the local environment of particle $i$ consists of an 8-dimensional vector, 
\begin{equation}
\label{input}
\mathbf{Q}(i) = (\{\bar{q}_l(i)\}),
\end{equation}
with $l\in[1,8]$.

The set of nearest neighbors of each particle is identified with a parameter-free criterion called SANN (solid angle nearest neighbor)\cite{SANN} for the hard spheres and Wahnstr\"om models. In this approach, an effective individual cutoff radius, $r_c(i)$, is found for every particle $i$ in the system based on its local environment. This method is not inherently symmetric, i.e. $j$ might be a neighbor of $i$ while $i$ is not a neighbor of $j$. However, symmetry can be enforced by either adding $j$ to the neighbors of $i$ or removing $i$ from the neighbors of $j$. In this study, we applied the latter solution. For the Kob-Andersen mixture, we obtained better results with a fixed cutoff radius (see SI for a comparison).

\subsection{Unsupervised machine learning}

The UML approach used here follows the method outlined in Ref. \cite{boattini2019unsupervised}. A detailed description is provided in the SI.


\end{document}